# Verification of A Security Adaptive Protocol Suite Using SPIN

Shamim Ripon, Sumaya Mahbub and K. M. Intiaz-ud-Din

*Abstract*—The advancement of mobile and wireless communication technologies in recent years introduced various adaptive protocols to adapt the need for secured communications. Security is a crucial success factor for any communication protocols, especially in mobile environment due to its ad hoc behavior. Formal verification plays an important role in development and application of safety critical systems. Formalized exhausted verification techniques to analyze the security and the safety properties of communications protocols increase and confirm the protocol confidence. SPIN is a powerful model checker that verifies the correctness of distributed communication models in a rigorous and automated fashion. This short paper proposes a SPIN based formal verification approach of a security adaptive protocol suite. The protocol suite includes a neighbor discovery mechanism and routing protocol. Both parts of the protocol suite are modeled into SPIN and exhaustively checked various temporal properties which ensure the applicability of the protocol suite in real-life applications.

## I. Introduction

A mobile ad hoc network is a collection of wireless mobile nodes that dynamically self-organize into arbitrary and temporary network topologies [1]. Due to the ad hoc properties of mobile wireless communication the scenario of communication has changed dramatically in recent years. In a mobile and wireless environment, the network infrastructure is not fixed and the network nodes are not only movable and their participation is ad hoc. In such a network, the odes can randomly move into the network organizing themselves arbitrarily and leave it at any point in time. Hence the network topology can change rapidly and unpredictably. A major constraint in the design of scalable ad hoc networks is the mobility of mobile nodes. The realistic movements of mobile nodes are captured by mobility models [2]. Mobile nodes that are within each other's range communicate directly via wireless links, while those that are far apart rely on other nodes to relay messages as routers [3]. Each node guides the routing messages according to the routing protocol designed for such kind of networks.

The routes and the flow of the packets in a mobile network are decided by an ad hoc routing protocol. The general mechanism of ad hoc routing protocols is that, message is broadcasted to discover that path from the source and the destination node. Data packets are then sent over that path. An ad hoc routing protocol could be proactive (table driven), reactive (on demand) or hybrid [4]. Table driven routing protocols maintains a table, consisting of routes to the destination, at the regular interval of time, by exchanging the table information between the nodes periodically. Reactive routing protocols on the other hand initiate the route discovery only when a node requires a route to the destination node, to which the node wants to send the data. That is why they are also termed as on-demand routing protocols. Examples of reactive routing protocols are AODV, Dynamic Source Routing (DSR), Temporally Ordered Routing Algorithm (TORA), Associativity-Based Routing (ABR), Signal Stability Routing (SSR) and many others. Hybrid routing protocols are the combination of both the techniques used in proactive and reactive routing protocols. The technique applied according to the demand of the situation. Examples of hybrid routing protocols are Core Extraction Distributed ad hoc Routing (CEDAR) protocol, Zone Routing Protocol (ZRP) and Zone based Hierarchical Link State (ZHLS) routing protocol.

Numerous approaches have been proposed to analyze the security and routing properties of ad hoc protocols. These techniques include visual inspection, network simulation, analytical proofs, simulatability models, and formal methods [5]. Formal method is one of the most reliable ways to rigorously check the desired properties of a protocol by modeling and analyzing it mathematically. This paper uses model checking approach to automatically evaluate the safety-critical properties of a security adaptive protocol suite.

Model checking is an automated technique where first, models of both system and properties are created, then the model checker checks whether the model satisfies the specified properties. Within appropriate constrains, a model checker can perform an exhaustive state-space search on a software design or implementation and alert the implementing organization to potential design deficiencies by producing a counter example. In this paper we use SPIN model checker [6] where models are specified by using PROEMLA language. Our primary aim is to demonstrate design flaws that lead to violations of security requirements using model checking. Model checkers are good at finding design errors and they provide error traces (i.e., counter-examples). In our earlier work we developed a security adaptive protocol suite [7] that consists of a neighbor discovery mechanism and a routine protocol. In this paper we model checks the security protocol to find errors in its formal specification.

In the rest of the paper, we give a brief review of our security adaptive protocol suite in Section II which is followed by an overview of SPIN model checker in Section III. Section IV illustrates the proposed framework for formal verification of the protocol suite. Finally, Section V concludes the paper and outlines our future plan.

## II. SECURITY ADAPTIVE PROTOCOL SUITE

The security adaptive protocol suite is a combination of a neighbor discovery mechanism and a protocol. In the neighbor discovery phase, the neighbor nodes are ranked which is considered the trustworthiness of the nodes. With the ranked information collected expressing the trust of the neighbors, the routing protocol proceeds. When a demand is made from a source to a destination, the source first judges the security requirement of the application and then based on the ranking information of the neighbor nodes the packet is routed to the destination.

### A. Ranked Neighbor Discovery (RND)

Estimated physical distance between nodes is used for ranking neighbors. If a node is physical further away than that in routing table, a wormhole is present in the network. The idea is to add some information to the packets that restricts the maximum allowed transmission distance.

The nodes are tightly synchronized with a clock and $\Delta t$ is the allowed difference between the nodes' clocks. Temporal leashes of a packet is authenticated by using TESLA with Instant Key-disclosure (TIK). TESLA combines the advantage of digital signature and Message Authentication Codes (MAC). TESLA requires that the MAC value of a packet is received earlier by the receiver than the time at which TESLA key is used to compute the MAC by the sender. This can be achieved by sending the MAC value at the beginning of the transmission and TESLA key at the end of the transmission, shown in Fig. 1.

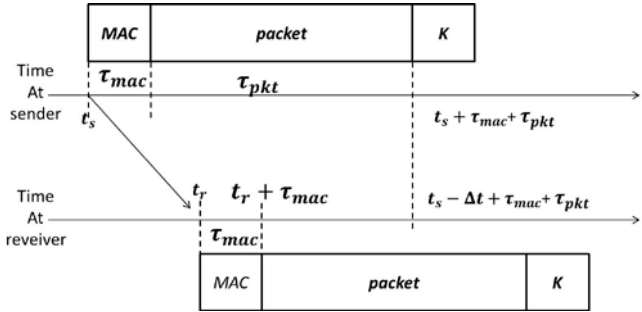

Figure 1 Synchronous timing diagram for sender and receiver

The receiver checks for the TESLA condition to be satisfied, which is that the MAC received before key is released and the receiver can start the verification of the MAC immediately

$$t_r + T_{mac} < t_s - \Delta t + T_{mac} + T_{pkt}.$$

After the synchronization of processes it is required to measure the distance. The receiver computes an upper bound on its distance $d'$ to the sender as follows:

$d' = V_{light}(t_r - t_s - \Delta t)$ [$V_{light}$ = speed of light]

Based on the calculated values the sender decides the trust values by ranking the neighbors based on predefined range. Four (4) types of ranking is used in our protocol suite (shown in Table 1).

Table 1 Rank assignment

| Distance Estimation | Rank (R) |
|---|---|
| $d' \leq T/4$ | 4 |
| $T/4 < d' \leq T/2$ | 3 |
| $T/2 < d' \leq 3T/4$ | 2 |
| $3T/4 < d' \leq T$ | 1 |
| $T > d'$ | 0 |

### B. Security Adaptive Ad-Hoc On-Demand Vector Routine

Ad hoc on demand distance vector (AODV) is a routing protocol for mobile ad-hoc networks. With RND algorithm it is now possible to identify the presence of wormholes in the network. Each node is assigned a ranked value. Whenever a route request is made from source to destination, the minimum security level (MSL) is checked with the assigned ranked values and transmit the message only to the trusted nodes. Refer to [7] for further reading.

## III. SPIN MODEL CHECKER OVERVIEW

SPIN [3] is an automata-based temporal logic model checker. Its specification and modeling language is called PROMELA. In PROMELA, a system is modeled as the composition of asynchronous processes that can interact with buffered or unbuffered message channels. Since there is no notion of time or clock, models with real-time aspects are very hard, if not impossible, to express in PROMELA. The language is especially designed to describe systems such as asynchronous communication protocols.

Correctness properties are specified in several ways in PROMELA. Assertions and *never* claims are the most frequently used constructs for this purpose. An assertion has a similar semantic as in the C Language. When the expression to be asserted is false, the assertion fails, and SPIN gives "assertion violated" error.

*Never* claims are used to specify the finite or infinite behavior that should never happen during the execution of a system. When we want to specify a property to be satisfied by the system, we formalize it in a logic formula and produce a *never* claim that corresponds to the negation of this formula. SPIN then tries to find a violation for this *never* claim. If it finds one, this means there is a case that the opposite of our property can occur in the system, which means our property can not be satisfied by the system. A *never* claim can be written by hand or can be translated from a linear temporal logic (LTL) formula. SPIN also includes a timeline property editor that helps users visually specify properties that are otherwise hard to formalize.

SPIN has two modes of operation: simulation and verification. In simulation mode, it runs the model and helps users get an impression on how their model behaves and debug their model. In verification mode, SPIN analyzes the model against the properties considering all possible executions performing an exhaustive search on the state space. It can also perform partial search on the state space, which is quite useful in case of very large models or insufficient computational resources.

If SPIN finds a violation, it produces an error trace. Using this error trace, a user can run a simulation of the execution that leads to the violation.

Our primary aim in this work is to find security flaws in the design of SAODV protocol. We employ SPIN to check our protocol model against the security properties that we formally specify as never claims in PROMELA and to list any security flaws, if any, as violations.

## IV. PROPOSED MODELING OF PROTOCOL SUITE

A key task in modeling is abstracting various unnecessary details from specifications. Some assumptions of the models need to be considered as well.

The nodes in our protocol suite are modeled using PROMELA process types. We start by modeling one source, one destination. Depending on the size of the network we fix the number of intermediate nodes for simplicity. Initially, we consider only a single attacker.

Broadcast communication among nodes are modeled by using PROMELA channels. In a network with $N$ nodes, each node is associated with $N-1$ unidirectional channels, and only the channels to its neighbors are used to send and receive messages. Separation of channels allows us to use channel assertions to reduce the size of the state space. The capacity of each channel is limited to two messages, which is enough for our modeling purposes. The neighborhood information is stored in a global connectivity matrix. Since the links are assumed to be bidirectional, the connectivity matrix is symmetric. When broadcasting, nodes refer this matrix to decide whether to send a message to a node or not.

Two important properties are considered initially for routing operation:

i) Correctness of distance information: We consider that if there is a route from a source to a destination, then the length of this path that is known to source is the shortest path between the two nodes. These properties can be specified by using temporal logic (LTL).

ii) Loop freedom: Checking the absence of cycle among the nodes is another important property to be considered during model checking. We can say that if there is a path from source to a destination then the number of hops between them is at most $n-1$ if there are n nodes in a network.

## V. CONCLUSION

In this short paper we have proposed a formal verification approach to check the security, safety and other critical properties of a security protocol suite. SPIN model checker is to be used for the purpose of automated model checking. LTL formulas can model the safety and security properties which can be verified against the model of the protocol. SPIN can manage to reveal the flaws in the protocol specification in a couple of seconds or even in a sub-second time scale. Importantly, SPIN draws a counterexample identifying the exact state transitions that introduce flaws in the specification.

We are very early stage of model checking the protocol suite. Apart from checking whether the neighbour discovering and routing of the protocol our future plan includes checking the security properties involving mobility and multiple attackers, causing secure routing a challenging task. In long term we plan to model check other similar routing protocols and draw a conclusive comparison among the protocols.